# Current-voltage characteristic, stability, and self-sustained current oscillations in resonant-tunneling n-doped semiconductor superlattices

L. L. Bonilla, M. Kindelan, M. Moscoso, and A. Wacker*
*Escuela Politécnica Superior, Universidad Carlos III de Madrid, Butarque 15, 28911 Leganés, Spain*

We review the statics and dynamics of electric-field domains on doped superlattices within a discrete drift model. A complete analysis of the construction and stability of stationary field profiles having two domains is carried out. As a consequence we determine the intervals of doping on which self-sustained current oscillations may appear under dc voltage bias. We have also studied the influence of doping, boundary condition and length of the superlattice on the self-sustained oscillations. Our study shows that there are bistability regions where either self-sustained current oscillations or steady states are reached depending on the initial condition. For a wide bias interval, the self-sustained oscillations are due to the formation, motion and recycling of electric-field domain walls inside the superlattice. There are biases (typically in the region of bistability) for which the strength of the high and low field domains changes periodically in time while the domain wall remains almost pinned on a few quantum wells.

## I. INTRODUCTION

Electrical transport in semiconductor superlattices (SL) has attracted much interest during the last years due to the very different interesting properties related to the artificial band structure. One of these features is the occurrence of stationary electric field domains which have been observed already in 1974 [1]. Due to advanced growing facilities and experimental techniques the complicated structure of the current-voltage characteristics which exhibits several branches, roughly equal to the number of quantum wells, could be resolved during the last decade [2–7]. In these experiments it was demonstrated that the different branches are connected to the formation of two domains with different values of the electric field inside the sample. Depending on the conditions, stable stationary field domains and traveling domain boundaries may occur. In the latter case, the dynamics of the electric field domains gives rise to time-dependent oscillations of the current. [8–11]

During the last years it has been shown theoretically that the observed phenomena can be fairly well reproduced by models which essentially combine the discrete Poisson-equation and rate equations for the carrier densities in the different quantum wells [12–15]. Also the time-dependent current oscillations could be recovered in these models [14,16,17]. A prediction of spatio-temporal chaos in resonant-tunneling superlattices under dc+ac voltage bias has been made on the basis of the discrete drift model [18]. The influence of growth-related imperfections on the SL behavior has been studied in [19]. These phenomena may also be described by discrete models with Monte Carlo dynamics incorporating single-electron tunneling effects (wich are important for slim superlattices and give rise to additional oscillations of the current [20,21]).

In this paper we want to explain *how* these complicated phenomena are generated by such models. This provides a deeper insight into the basic mechanisms and helps to classify the results of various experiments and computer simulations. In particular we want to give an answer to the following questions: How is it possible to understand the appearance of the complicated structure of the current-voltage characteristic? What are the conditions for stability and oscillations and how can they be understood? What is the main difference to the Gunn diode, where hardly any stable domain states are observed? Where can the high-field domain be located with respect to the injecting contact? What are the mechanism(s) explaining the current oscillations? How do doping, electron velocity and boundary conditions influence the shape and frequency of the current oscillations?

The paper is organized as follows: The model we use is described in the second section. The third section shows how the complex stationary current-voltage characteristic changes as the doping increases. In the fourth section we investigate the stability of the stationary states and prove an explicit criterion for the occurrence of stable domain states. The fifth section provides a thorough numerical investigation of the oscillatory behavior generated by the instability of the current discussed before. The last section contains our conclusions and the Appendix is devoted to a proof that no self-sustained oscillations appear for low enough doping times number of SL periods.

## II. THE MODEL

We consider a semiconductor superlattice where the lateral extension of wells and barriers is much larger than the total length of the SL, so that single-electron tunneling effects (see e.g. Ref. [20]) are negligible. The quantum wells (QW) are weakly coupled and the scattering times are much shorter than the tunneling time between adjacent QWs. Thus it makes sense to consider the electrons to be localized within the QWs and in local equilibrium at the lattice temperature. The current is mainly determined by the resonances between the different energy levels in the QWs, which we denote by Ci, $i = 1, 2, \ldots$, in order of increasing energy counted from the bottom of the conduction band. For the biases of interest here, there are three important resonances C1C1, C1C2 and C1C3. If the intersubband relaxation is also fast with respect to the tunneling, in practice only the lowest subband is occupied. In this case it is a reasonable approximation to consider the QWs as entities characterized by average values of the electron density $\tilde{n}_i$ in the $i$-th QW and the electric field $\tilde{E}_i$ between wells $i$ and $i+1$, with $i = 1, \ldots N$. The discrete drift model consists of the following system of equations for $\tilde{E}_i$, $\tilde{n}_i$, and the total current density $\tilde{J}(\tilde{t})$ [16]:

$$\tilde{E}_i - \tilde{E}_{i-1} = \frac{q\,\tilde{l}}{\epsilon}\,(\tilde{n}_i - \tilde{N}_D), \tag{1}$$

$$\epsilon\,\frac{d\tilde{E}_i}{d\tilde{t}} + q\,\tilde{v}(\tilde{E}_i)\,\tilde{n}_i = \tilde{J}, \tag{2}$$

$$\tilde{l}\sum_{i=1}^{N}\tilde{E}_i = \tilde{\Phi}, \tag{3}$$

where $i = 1, \ldots, N$. In this model Eq. (1) and Eq. (2) are, respectively, the one-dimensional Poisson equation (averaged over one SL period) and Ampère's law. $\epsilon$, $\tilde{N}_D$, $\tilde{l}$ and $q$ are the average permittivity, average donor concentration, SL period, and the charge of the electron. $\tilde{v}(\tilde{E})$ is an effective electron velocity that has a peak at certain values of the electric field connected to resonant tunneling C1C1, C1C2 and C1C3. In this paper we shall be concerned with phenomena occurring at fields higher than the first resonant peak C1C2, so that we shall omit the miniband peak C1C1 in our velocity curve; see the curve plotted in Fig. 1 by a dashed line. Phenomena at lower fields can easily be studied by adding the C1C1 peak to our curve, [23]. Equation (3) establishes that the average electric field is given by the *dc* voltage bias $\tilde{\Phi}$. Notice that there are 2N+2 unknowns: $\tilde{E}_0, \tilde{E}_1, \ldots, \tilde{E}_N, \tilde{n}_1, \ldots, \tilde{n}_N, \tilde{J}$ and 2N+1 equations so that we need to specify one boundary condition for $\tilde{E}_0$ plus an appropriate initial profile $\tilde{E}_i(0)$. The boundary condition for $\tilde{E}_0$ (the average electric field *before* the SL) can be fixed by specifying the electron density at the first site, $\tilde{n}_1$, according to (1). In typical experiments the region before the SL has an excess of electrons due to a stronger n-doping there than in the SL [8–11]. Thus it is plausible assuming that there is an excess number of electrons at the first SL period measured by a dimensionless parameter $c > -1$:

$$\tilde{n}_1 = (1 + c)\tilde{N}_D\,. \tag{4}$$

$c$ has to be quite small because it is known that a steady uniform-electric-field profile is observed at low laser illumination in undoped SL [8,14,16]. This observation allows us to infer the electron velocity directly from measured current-voltage data [14]. Another possibility is to derive the electron velocity from simple one-dimensional quantum-mechanical calculations of resonant tunneling, as was done by Prengel et al [13]. They used a more complicated discrete model with two electron densities corresponding to the populations of the two lower energy levels of each QW. Their model reduces to a form of ours when the large separation between the time scales of phonon scattering, resonant tunneling and dielectric relaxation is taken into account. An earlier discrete model is due to Laikhtman [24] and an attempt at deriving a discrete model from quantum kinetics can be found in [12,15] (only miniband conduction was considered, not resonant tunneling).

For the calculations that follow, it is convenient to render the equations (1)-(4) dimensionless by adopting as the units of electric field and velocity the values at the C1C2 peak of the velocity curve, $\tilde{v}(\tilde{E})$, $\tilde{E}_M$ and $\tilde{v}_M$ (about $10^5$V/cm and 427 cm/s, respectively, for the sample of Ref. [10]). We set [16]:

$$E_i = \frac{\tilde{E}_i}{\tilde{E}_M}\,, \qquad n_j = \frac{q\tilde{l}\tilde{n}_i}{\epsilon\tilde{E}_M}\,, \qquad I = \frac{\tilde{J}}{qN_D\,\tilde{v}_M}\,,$$
$$v = \frac{\tilde{v}}{\tilde{v}_M}\,, \qquad t = \frac{qN_D\,\tilde{v}_M\,\tilde{t}}{\epsilon\tilde{E}_M}\,, \qquad \phi = \frac{\tilde{\Phi}}{N\,\tilde{E}_M\,\tilde{l}} \tag{5}$$

By using Poisson's equation to eliminate the electron density in favor of the electric field we obtain the dimensionless equations:

$$\frac{dE_i(t)}{dt} = I(t) - \left(1 + \frac{E_i - E_{i-1}}{\nu}\right)\,v(E_i) \tag{6}$$

for $i = 1, \ldots N$,

$$\phi = \frac{1}{N}\sum_{i=1}^{N} E_i(t)\,, \tag{7}$$

and the boundary condition

$$E_0(t) = E_1(t) - c\nu. \tag{8}$$

The dimensionless parameter $\nu$, is defined by

$$\nu = \frac{\tilde{N}_D\,q\,\tilde{l}}{\epsilon\,\tilde{E}_M}, \tag{9}$$

which yields about 0.1 for the SL used in the experiments [10,11]). The constant voltage condition (7) determines the current to be

$$I(t) = \frac{1}{N}\sum_{i=1}^{N}\left(1 + \frac{E_i - E_{i-1}}{\nu}\right)\,v(E_i) \tag{10}$$

With the choice (5), the dimensionless velocity $v(E)$ has a maximum at $E = 1$ with $v(1) = 1$. Throughout this paper we use the function $v(E)$ which is plotted in Fig. 1 by a dashed line. Besides having a maximum $v(1) = 1$, it has a minimum at $E_m \approx 1.667$ with $v(E_m) = v_m \approx 0.323$. Nevertheless nearly all of the features discussed in the following are independent of the exact shape of the function $v(E)$. We only impose the restrictions that $v(E) > 0$ for $E > 0$ and the existence of a minimum at $E_m > 1$.

For any $\phi > 0$, we shall assume that the initial electric field profile is strictly positive and that the electron density is non-negative: $E_i(0) > 0$, $n_i(0) \equiv E_i(0) - E_{i-1}(0) + \nu \geq 0$, $\forall i$. This is reasonable unless $\phi$ is close to zero (but then the C1C1 peak in the velocity curve should be restored) or the boundary conditions are unrealistic. From the equations and our assumption on the initial field profile, it follows that $E_i > 0$ and $n_i \equiv E_i - E_{i-1} + \nu \geq 0$ for all positive times. The model equations have interesting properties concerning the monotonic behavior of the electric fields with respect to the QW number. These are summarized in the following Lemmas:

**Lemma 1** *If the fields of two adjacent QWs are identical, i.e.,* $E_k = E_{k-1}$ *holds for some* $k$ *with* $2 \leq k \leq N$, *there is at least one* $i$ *from* $1 \leq i \leq k$ *with* $E_i = E_{i-1}$ *and* $d(E_i - E_{i-1})/dt \neq 0$. *For* $c = 0$ *there is the additional possibility that* $E_0 = E_1 = \ldots = E_k$ *holds.*

**Lemma 2** *If* $c \geq 0$ *and the field distribution is monotone increasing at* $t = 0$ *(*$E_i(0) \geq E_{i-1}(0), \forall i$*), it will continue being monotone increasing with respect to the well index* $i$ *for all later times* $t > 0$. *(For* $c \leq 0$, *the same holds for a monotone decreasing field distribution).*

Both lemmas can be easily proved by using the property

$$\frac{d}{dt}(E_k - E_{k-1}) = \frac{E_k - E_{k-1}}{\nu}\, v(E_{k-1}), \tag{11}$$

which holds for the particular $k$ considered in the first Lemma, and the boundary condition (8).

## III. STATIONARY STATES

In this chapter we want to explain how the complex domain structures found experimentally and from computer simulations [5,7,13,14] are generated by this simple model.

We denote the electric field profile and the current density of stationary states by $E_i^*$ and $I^*$, respectively. An easy way to construct the stationary profiles is to fix $I^*$, find out the corresponding electric field profile $\{E_i^*\}$, $i = 1, \ldots, N$, calculate their voltage as a function of $I^*$

$$\phi(I^*) = \frac{1}{N} \sum_{i=1}^{N} E_i^*(I^*)\,. \tag{12}$$

The field profile must fulfil the equation [14]:

$$E_{i-1}^* = E_i^* + \nu \left(1 - \frac{I^*}{v(E_i^*)}\right) =: f(E_i^*, I^*) \tag{13}$$

The boundary condition implies

$$v(E_1^*) = \frac{I^*}{c+1} \Leftrightarrow f(E_1^*, I^*) = E_1^* - c\nu \tag{14}$$

which has three solutions $E_1^*$ for a known fixed value of the current on the interval $(1+c)\, v_m < I^* < 1 + c$.

In order to understand the properties of the stationary profiles we will now investigate the behavior of the set $\{E_i^*\}$ as a function of $E_1^*$. At first we restrict ourselves to $c > 0$ and monotone increasing field profiles. To construct $\{E_i^*\}$, we have to invert the function $f(E, I^*)$ for a fixed value of $I^*$. Its derivative is:

$$\frac{\partial f(E, I^*)}{\partial E} = 1 + \nu \frac{I^*}{v(E)^2} \frac{dv(E)}{dE} \tag{15}$$

With the restriction to monotone increasing field profiles, we can always obtain $E_i^*$ for $E_1^* \geq E_m$ because $dv/dE > 0$ holds. If $E_1^* < E_m$, Eq. (14) implies $I^* \leq 1 + c$ and we find that $f(E, I^*)$ is strictly monotone increasing for all $E$ if

$$\frac{1}{\nu} \geq \max\left\{ \frac{-(1+c)}{v(E)^2} \frac{dv(E)}{dE} \right\} \tag{16}$$

For our function $v(E)$ this yields $\nu \leq 0.195/(1+c)$. In this case the function $f(E, I^*)$ ($I^*$ fixed) is always invertible. Then we can find a unique field profile parametrized by the point $E_1^*$. Since we have three possible solutions $E_1^*$ of (14) for each given $I^*/(1+c) \in (v_m, 1)$, there are three different voltages $\phi$ for each value of the current in this range. The function $\phi(I^*)$ is thus three-valued, which means that by inverting it we obtain an N- or an Z-shaped current-voltage characteristic as shown in Fig. 1 for $\nu = 0.05$ and $\nu = 0.15$, respectively. Both types can be easily understood: When the doping density $\nu$ is low, Eq. (13) shows that $E_i^* \approx E_1^*$ holds. Thus, the field profile is nearly uniform and the current-voltage characteristics follows the $v(E)$-curve as shown in Fig. 1 for $\nu = 0.05$. This is physically obvious as there are few charges present inside the sample. For larger values of $\nu$ the values $E_i^*$ may strongly deviate from $E_1^*$ with increasing $i$, if $E_1^*$ is not a fixed point of $f(E, I^*)$ (which is the case for $c = 0$ [14]). Let us denote by $E^{(1)}(I) < E^{(2)}(I) < E^{(3)}(I)$ the three solutions of $v(E) = I$ for a given $I \in (v_m, 1)$ which are the fixed points of Eq. (13). If $\phi$ is small and $c > 0$, $E_1^*$ is on the first branch of $v(E)$ and the values $E_i^*$ tend to $E^{(1)}(I^*)$. When $\phi$ is larger and $E_1^*$ is located on the second branch of $v(E)$, the sequence $E_i^*$ leaves the neighborhood of $E^{(2)}(I^*)$ and then approaches $E^{(3)}(I^*)$ on the third branch of $v(E)$ if $\nu$ is large enough. This is shown in Fig. 2(a). In this case the voltage is basicaly determined by the fixed point $E^{(3)}(I^*)$. Since $dE^{(3)}(I)/dI > 0$, this branch of stationary solutions may exhibit a range of positive differential conductance leading to the Z-shape. This effect is more pronounced for longer superlattices with many wells $N$ and also for larger values of $c$.

For larger doping $\nu$ the condition (16) is violated and the function $f(E, I)$ may not be invertible for some current $I$. In this case there can be more than one possible $E_{i+1}$ following a given $E_i$. Then the current-voltage characteristic can no longer be unambiguously parametrized by the point $E_1^*$. In general $f(E, I)$ has 3 different branches for a certain interval of $I^*$, as shown in Fig. 2(b). Let us call branch $\alpha$ that having $\partial f/\partial E > 0$ for low $E$, branch $\beta$ has $\partial f/\partial E < 0$, and branch $\gamma$ again has $\partial f/\partial E > 0$ but for larger $E$.

Let us explain how to construct different stationary field profiles for a given value of the current $I^*$. We shall assume that the profiles are monotone increasing, $E_{i+1}^* \geq E_i^*$, $\forall i$. First of all, $E_1^*$ may be located on branch $\gamma$ of $f(E, I^*)$, and so will be all successive fields $E_i^*$. This profile will have the largest possible voltage for the same $I^*$. Secondly, $E_1^*$ may be located on branch $\beta$, which implies that all successive $E_i^*$ of a monotone increasing field profile have to be on branch $\gamma$. The corresponding voltage is smaller than that of the previously described branch but larger than those stationary solution branches that we analyze next.

If $E_1^*$ is on branch $\alpha$, we may have $E_i^*$, $i = 1, \ldots, j-1$, $(j = 2, \ldots, N)$ on branch $\alpha$, and $E_j^*$ either on branch $\beta$ or on branch $\gamma$. We obtain a different branch of stationary solutions for each such possibility. Let us denote by $(j, \beta)$ or $(j, \gamma)$ the solution branch having $E_j^*$ either on branch $\beta$ or $\gamma$ respectively, and $E_i^*$, $i = 1, \ldots, j-1$ on branch $\alpha$. In Fig. 2(b) a solution $(j, \beta)$ is shown by a dashed line and a solution $(j, \gamma)$ by a full line. Clearly $j = 1$ corresponds to the possibilities discussed above. Finally, we have one solution where all field values are on branch $\alpha$ which we denote by $(N+1, \gamma)$. In order of increasing voltages, we have

$$\phi_{(N+1,\gamma)}(I^*) \leq \phi_{(N,\beta)}(I^*) \leq \phi_{(N,\gamma)}(I^*) \leq \\ \ldots \leq \phi_{(1,\beta)}(I^*) \leq \phi_{(1,\gamma)}(I^*),$$

corresponding to $2N+1$ different stationary solution branches with the same current $I^*$. They can be observed in Fig. 1 for $\nu = 1.0$ and $I^* = 0.8$. Notice that the branches $(j+1, \gamma)$ and $(j, \beta)$ coalesce at a current $I^* \in (1, 1+c)$ which is roughly independent of $\nu$ (see Fig. 2(c)). The branches $(j, \beta)$ and $(j, \gamma)$ coalesce at a lower current $I_c$ which decreases as $\nu$ increases (see Fig. 2(d)). The current-voltage characteristic curve is thus connected as shown in Fig. 1 ($\nu = 1.0$).

The field profile of the solution branch $(15, \gamma)$ is depicted by the crosses in Fig. 3. One can clearly identify two regions $1 \leq i < j$ and $j < i \leq N$ where the electric field $E_i^*$ is roughly constant and close to a fixed point with $v(E_i^*) \approx I^*$. In between there is a transition layer, the domain boundary, consisting of only a few wells. These type of states we call domain states. A shift of the domain boundary by one well only changes the voltage as long as the transition layer does not extend to one of the contacts. As the stationary solutions resulting from a one-well shift are very similar, the domain branches in the current–voltage characteristics look alike, as can be seen in Fig. 1 ($\nu = 1.0$). The slope of the different domain branches may vary with the position of the domain boundary due to the different conductivity of the two domains which is related to the shape of the $v(E)$ curve.

Note that the domain states are not very sensitive to the exact type of boundary conditions if two conditions are fulfilled:

1. The boundary conditions must allow for the existence of a roughly constant field distribution $E_i^* \approx E^{(1)}$ and $E_i^* \approx E^{(3)}$ probably after a short contact layer of some wells.

2. The domain boundary must be located sufficiently deep inside the sample, so that it does not collide with the contact layer.

As $\nu$ decreases, the solution branches become shorter and eventually disappear if $I_c$ becomes larger than $(1+c)$, which happens if the inequality (16) holds. The stationary domain structures are seen for narrow current intervals about $I^* = 1$ for intermediate doping as shown in Fig. 1($\nu = 0.3$). Another complex feature can be found here for larger voltages where extra wiggles appear. They occur if $E_1^*$ crosses the value 1, yielding an additional maximum in $I^*$.

Except for the fundamental question of stability, we have now understood the morphology of the complicated current-voltage characteristic curve shown in Fig. 1, its changes with doping and its relation with the electric field profile. The very same features occur in the more complicated model of Prengel et al. [13,17] as shown numerically in Ref. [25].

So far we have restricted ourselves to monotone increasing field profiles. Therefore we could only find domain states where the high-field domain is located at the receiving contact and the domain boundary is an accumulation layer. Nevertheless, this is not the full story. For sufficiently large $\nu$ other solutions are also possible, even for $c > 0$. A typical such field profile is depicted by circles in Fig. 3. The field starts on branch $\gamma$ of the function $f(E, I^*)$ and first increases with the QW index towards the third fixed point $E^{(3)}(I^*)$. At a certain QW $j$ the field jumps down to either branches $\alpha$ or $\beta$, and then decreases down towards the first fixed point $E^{(1)}(I^*)$. Of course $\nu$ has to be large enough for these jumps down to be possible. Thus these field profiles have a high-field domain located at the injecting contact and the domain boundary separating this domain from a low-field domain is a depletion layer. Numerical investigation shows that these stationary states are stable and that they can be reached from many initial conditions. Lemma 2 tells us that the initial field distribution can not be monotone increasing in the well index (like the solutions of the connected branch discussed before), for otherwise the field distribution would stay monotone increasing for all times. Thus, these different solutions are not connected to the branches dicussed before (having a monotone increasing field profile) but form many additional isolated curves called isolas [26]. A typical isola is shown inside the frame in Fig. 1 for $\nu = 1.0$, which is also blown up to an enlarged scale for the sake of clarity.

A special situation arises if $c = 0$ holds. In this case the branches of monotone increasing and decreasing field profiles become connected and there appears much additional degeneracy leading to an extremely complicated structure as observed in Ref. [14].

## IV. STABILITY OF THE STATIONARY STATES

Up to now we have only discussed the existence of stationary states, but not their stability properties. First of all, several stability properties can be established by topological arguments [26,27]. If several branches overlap at a fixed voltage, each second branch has to be unstable by general reasons. For example the middle branch (exhibiting positive differential conductivity) of the Z-shaped charactistics in Fig. 1($\nu = 0.15$) has to be unstable. The remaining branches may exhibit further bifurcations. In order to elucidate this, we will perform a linear stability analysis for the states constructed in the previous section. By this method we will prove the following statements:

A. For large doping $\tilde{N}_D$ exceeding approximately $\epsilon(\tilde{E_m} - \tilde{E_M})/(q\tilde{l}) \cdot \tilde{v_m}/(\tilde{v_M} - \tilde{v_m})$ we find stable domain states.

B. For very small products $\tilde{N}_D(N-1)$ the almost uniform states are also stable.

For a medium range of doping in between these two limits we find self generated oscillations as reported in Ref. [10,11], and further discussed in the Section V.

In order to perform the linear analysis of stability, we set

$$E_i(t) = E_i^* + e^{\lambda t}\,\hat{e}_i \tag{17}$$

$$I(t) = I^* + e^{\lambda t}\,\hat{j} \tag{18}$$

and obtain

$$\lambda \hat{e}_i = \hat{j} - \frac{I^* v'(E_i^*)}{v(E_i^*)}\hat{e}_i - \frac{v(E_i^*)}{\nu}(\hat{e}_i - \hat{e}_{i-1}) \tag{19}$$

This linear equation together with the boundary condition $\hat{e}_0 = \hat{e}_1$ determines all $\hat{e}_i$ as a function of $\lambda$. The fixed bias condition

$$\sum_{i=1}^{N} \hat{e}_i = 0, \tag{20}$$

then determines the possible eigenvalues $\lambda$. For $\lambda \neq 0$ we now introduce the variable $Y_i = \lambda\,\hat{e}_i/\hat{j}$ and the parameters

$$b_i = I^* \frac{v'(E_i^*)}{v(E_i^*)}, \quad a_i = \frac{v(E_i^*)}{\nu}. \tag{21}$$

Then we obtain:

$$Y_i = \frac{\lambda + a_i Y_{i-1}}{\lambda + a_i + b_i} \quad \text{or} \quad \left\{ \begin{array}{c} \lambda + a_i + b_i = 0 \\ \text{and} \quad Y_{i-1} = -\lambda/a_i \end{array} \right\} \tag{22}$$

$$Y_1 = \frac{\lambda}{\lambda + b_1} \tag{23}$$

**Lemma 3** *If $b_i \geq 0$ holds for all $i = 1, \ldots N$ the real parts of all eigenvalues have to be negative, i.e., the state is stable. Furthermore the current-voltage characteristic exhibits a positive slope $dI^*/d\Phi$.*

We prove this Lemma by contradiction. Let us assume that $Re(\lambda) \geq 0$ holds with $\lambda \neq 0$. As $b_1 \geq 0$ we directly find that $Re(Y_1) > 0$ and $|Y_1 - 1| \leq 1$. In order to satisfy the voltage condition (20) we conclude that there must be at least one $Y_j$ with $Re(Y_j) < 0$. This implies directly that also $|Y_j - 1| > 1$ must hold. But we find:

$$|Y_i - 1| = \frac{|a_i(Y_{i-1} - 1) - b_i|}{|\lambda + b_i + a_i|} < \frac{|a_i||Y_{i-1} - 1| + |b_i|}{|b_i + a_i|} \tag{24}$$

Given that $|Y_1 - 1| \leq 1$, the last equation implies that $|Y_i - 1| \leq 1$ for all $i$. Thus, the case $Re(\lambda) \geq 0$, $\lambda \neq 0$ is excluded.

For $\lambda = 0$ we obtain

$$b_1 \hat{e}_1 = \hat{j} \qquad \text{and} \qquad (b_i + a_i)\hat{e}_i = \hat{j} + a_i \hat{e}_{i-1}. \tag{25}$$

Therefore all $\hat{e}_i$ have the same phase as $\hat{j}$ and the voltage condition (20) cannot be satisfied unless $\hat{e}_i = 0, \forall i$ and $\hat{j} = 0$, which is the trivial case. In conlusion $\lambda = 0$ is not an eigenvalue. Furthermore $\lambda = 0$ describes the infinitesimal change along the curve of stationary states. Eq. (25) tells us, that $\hat{j}/\hat{e}_i \geq 0, \forall i$. Identifying $dI^* = \hat{j}$ and $d\phi = \sum \hat{e}_i$ we obtain a positive slope of the current-voltage characteristic, i.e., $dI^*/d\phi \geq 0$.•

Therefore we can conclude that the states having all their fields $E_i$ in the positive differential mobility region are stable, which coincides with our physical intuition.

### A. Stability for sufficiently large $\nu$

Lemma 3 establishes that stationary field profiles with $v'(E_i^*) \geq 0$, $\forall i$ are linearly stable. These profiles include: (i) trivial ones where all the $E_i$ belong to the same branch of $v(E)$, and (ii) profiles where the negative differential mobility region is crossed in a single jump. This means that for a certain value $E_j^* \leq 1$ $(1 \leq j \leq N-1)$ there exists $E_{j+1}^* \geq E_m$ with $f(E_{j+1}^*, I^*) = E_j^*$. As $f(E, I^*)$ is monotone increasing for $E \geq E_m$, the necessary and sufficient condition for the existence of such a value $E_{j+1}^*$ is $f(E_m, I^*) \leq E_j^*$. This condition yields

$$(E_m - E_j^*)v(E_m) \leq \nu(I^* - v_m) \tag{26}$$

This inequality is first fulfilled for $E_j^* = 1$ and $I^* = 1$, which are the largest values of the respective quantities for the low field domain. This gives:

$$\nu \geq \frac{(E_m - 1)\, v_m}{1 - v_m} \tag{27}$$

If this condition is fulfilled, domain states are possible which cross the negative differential mobility region in a single jump and must be stable therefore.

Nevertheless there can be stable states even for smaller doping $\nu$, as Lemma 3 only yields a sufficient and not a necessary condition for stability. For our $v(E)$ curve inequality (27) yields $\nu \geq 0.32$. Indeed we do not find any self-sustained oscillations for $\nu$ larger than the value $\nu \approx 0.27$, which is somewhat smaller than our estimation. Checking other $v(E)$ curves we have always found oscillations up to a doping roughly $15-40\%$ lower than determined by the bound (27). Thus, the bound is not only a sufficient condition but also a reliable rough estimate for doping above which the oscillations disappear.

Transforming to physical units we obtain a surface charge density per well:

$$\tilde{l}\tilde{N}_D \geq \frac{\tilde{v_m}\, \epsilon\, (\tilde{E_m} - \tilde{E_M})}{(\tilde{v_M} - \tilde{v_m})\, q}\,, \tag{28}$$

which should be a reasonable approximation for the necessary doping density. For the model equations from Prengel [13] we find $\tilde{l}\tilde{N}_D \geq 2.4 \times 10^{11}/cm^2$. Actually, oscillations are found in the model up to roughly $\tilde{l}\tilde{N}_D \approx 10^{11}/cm^2$ for the regular superlattice and for somewhat higher values for a slight amount of disorder [19].

If we regard domains with the high-field domain located at the injecting contact the same arguments yield a bound

$$\nu \geq \frac{(E_m - 1)}{1 - v_m} \tag{29}$$

for the existence of stable domains which is larger by a factor $1/v_m$. For our $v(E)$ curve this corresponds to $\nu \geq 0.97$. Indeed we have found stable domains with an depletion layer for $\nu = 1.0$ as depicted in Fig. 3. This indicates that this type of domain only appears for larger doping. This might explain that two different locations of the high-field domain have been reported in the literature. In Ref. [4] it is found to be located at the injecting contact for a superlattice with $\tilde{l}\tilde{N}_D = 8.75 \times 10^{11}$ cm$^{-2}$ while in Ref. [28] the high-field domain is located at the receiving contact for a different superlattice with $\tilde{l}\tilde{N}_D = 1.5 \times 10^{11}$ cm$^{-2}$.

### B. Stability for sufficiently small $\nu$

Now we want to show that for sufficiently small doping, the (connected) branches of stationary solutions are stable. Then no self-sustained oscillatory branches bifurcating from them can exist. In order to do this, we note that for very low voltages the stationary state is stable as indicated by Lemma 3: all field values of this state are in the range $0 < E_i^* < 1$. We now increase the voltage and study whether an instability may occur by checking whether it is possible to have $\lambda = i\omega$ with $\omega > 0$ for some $\phi$. (The case $\lambda = 0$ yields the saddle-node bifurcation at the point with $d\phi^*/dI^* = 0$, which causes the switching to another branch of the Z-shaped characteristic but typically does not generate any oscillatory behavior.) In the appendix we show that this is possible only if

$$(N-1)\nu > \min\left\{ \frac{\pi\,(v_l - \nu c_1)}{4c_1}\,,\; \frac{v_l\, c_1}{2C\,(I^* - v_l)c_2} \right\} \tag{30}$$

holds, with

$$\begin{aligned} v_l &= \min\left\{ v_m\,,\, (1+c)^{-1} \right\} \\ c_1 &= I^* \max\nolimits_{E_l \leq E \leq E_h} |\partial \ln v(E)/\partial E| \\ c_2 &= I^* \max\nolimits_{E_l \leq E \leq E_h} |\partial^2 \ln v(E)/\partial E^2| \\ C &= \frac{v_l}{(N-1)\nu c_1}\left[ \exp\left( \frac{(N-1)\nu c_1}{v_l - \nu c_1} \right) - 1 \right]. \end{aligned}$$

Here $E_l, E_h$ denote the minimal and maximal values of the field for the stationary field profile. Note that for small $\nu$ we find $C \to v_l/(v_l - \nu c_1)$ and furthermore the terms $\nu c_1$ become neglegtible, so that the right side neither depends on $\nu$ nor on $N$ but only on the shape of $v(E)$ and the parameter $c$.

If $\nu$ is smaller and the inequality (30) is violated, no bifurcating oscillatory branches can issue forth from the steady state which is thereby stable.

The bound (30) is far too small due to the rough estimations made during its derivation. Therefore the number itself should not be used for quantitative investigations. Nevertheless we now have shown that the stationary states are stable for low doping and that in the limit of long superlattices the critical doping decreases as $1/(N-1)$.

### C. Consequences for the continuum limit

With respect to the continuum limit, $N \to \infty$, $\nu \to 0$, $L := N\nu < \infty$ we directly find that there exists a minimal length $L_m$ such that the stationary state is linearly stable if $L < L_m$. This lower bound is given by the Eq. (30) with $\nu = 0$, $C = 1$ and it can be derived directly from the equations valid in the continuum limit, as we shall report elsewhere [22]. Eq. (30) and similar bounds derived for other boundary conditions constitute an *explicit* form of the well-known $N_D L$ criterion of the Gunn effect [29]: The dimensionless length $L$ (proportional to doping times the semiconductor length [30]) has to be larger than a certain number for the stationary solution to be unstable.

Obviously the upper bound in the doping $\nu$ (for the absence of the oscillatory regime) does not exist in the continuum limit. The discreteness is essential for the field distribution to jump from the low-field region to the high field region without any fields exhibiting negative differential velocity in between, (which stabilizes the field distribution). This explains that these stable stationary domains can not be found in the usual Gunn diode.

## V. SELF-SUSTAINED OSCILLATIONS IN FINITE SUPERLATTICES

Here we report the results of numerical simulations of the self-sustained oscillations of the current in the discrete model for different values of the dimensionless parameters. We have solved Equations (6) and (7), with the boundary condition (8) and an appropriate initial field profile.

The simulations yield self-sustained current-oscillations for dopings $\nu$ where the stationary field profile becomes inhomogeneous and the middle branch exhibits currents that are significantly larger than $v(\phi)$. E.g., for the situation from Fig. 1 the oscillations are found for dopings $\nu > 0.1$. Then the middle branch generates oscillatory behavior in a certain range $[\phi_\alpha, \phi_\omega]$. We find $\phi_\alpha > 1$ and that $\phi_\omega$ is an increasing function of $c$ (due to the more pronounced Z-shape as discussed in the third section) which can become larger than $E_m$. In the latter case bistability between the oscillatory behavior and the lower branch of the Z-shaped characteristic, which is stable for $\phi > \phi_\beta \approx E_m$, occurs if $\phi_\omega > \phi_\beta$ holds. A similar type of bistabilty can be found, when the wiggles in the static $I^*(\phi)$ characteristic appear (Fig. 1, $\nu = 0.3$) and the upper branch is unstable against oscillatory behavior while the lower branch is stable. As discussed in the last section the oscillations completely vanish for dopings where the domain branches are fully developed.

To clarify the nature of the oscilations Fig. 4(a) shows the field profiles at different times of a given period of the current oscillations for a 50-well SL. We can identify a field profile consisting of two field domains at time (1). The domain wall, which is a charge monopole containing an excess charge of electrons ($n_i > 1$), moves towards the receiving contact in time (2-4) where it disappears. During this process a new domain wall emerges slowly as can be seen at the times (3) and (4), so that this process is repeated in time. Note that two different monopoles are present at the same time in a certain part of the period. Between the domain walls regions with a more or less homogeneous electric field are observed. We define the field of the domain to be the electric field $E_i$ at the position where $n_i$ has a local minimum, i.e., the variation in the field is minimal. In Fig. 4(b) these fields of the different domains are depicted by full lines as a function of time. Additionally we have shown the values of $E^{(1)}(I) < E^{(2)}(I) < E^{(3)}(I)$ (which are the three solutions of the equation $v(E) = I(t)$) as a function of time (dashed lines in the figures. Note that the solutions $E^{(1)}$ and $E^{(2)}$ disappear for $I > 1$). We find that the fields of the domains mainly follow the values $E^{(1)}(I)$ and $E^{(3)}(I)$. The difference corresponds to the displacement current. Monopole recycling and motion is clearer for longer SLs as shown in Fig. 5. Notice that now the field on the domains follows closely the dashed lines, as the displacement current is smaller due to the lower frequency.

The mechanism of the oscillation can be understood as follows: Let us start at the time marked by (1) in Fig. 5. At this instant we assume that the field profile consists of a low field domain with $E_l \approx E^{(1)}(I)$ next to the beginning of the SL and then a high field domain with $E_h \approx E^{(3)}(I)$. Within the low field domain there is a tiny inhomogeneity (contact layer) close to the injecting contact, which is due to the boundary condition. The constant voltage condition implies

$$E_l(t) = E_h(t) - \frac{N}{j(t)}\left(E_h(t) - \phi\right) \qquad (31)$$

The position $j(t)$ of the domain wall moves to the right with a certain velocity $v_{mon}/\nu$, where $v_{mon}$ is always significantly less than 1. Eq. (31) implies that $E_l$, $E_h$, and $I$ must increase with time so as to fulfil the fixed voltage condition. As long as the low field domain is in the region of positive differential velocity ($E_l < 1$), the field profile is stable and the contact layer follows adiabatically the current. At $t_0 \approx 129$ the field $E_l$ becomes larger than 1 and the contact layer loses its stability and injects charge into the sample. This charge accumulation $\delta n$ travels with a velocity of the order of $v(E_l)/\nu \approx 1/\nu$ and is amplified in time via

$$\frac{d\delta n(t)}{dt} = -v'(E_l)\,\delta n(t) \qquad (32)$$

with an initial value $\delta n(t_0) \sim c$. At time (3), $t_0 + T_f$, its size is sufficiently large so that two different field domains on both sides can be identified and a new charge monopole is born. This charge monopole travels than with the velocity $v_{mon}/\nu$ mentioned above and it sharpens as it travels; see the point marked by (4) in the figure. This stage lasts until the fields before and after the monopole reach the values $E = E^{(1)}(I)$ and $E = E^{(3)}(I)$, respectively, and we are back in situation (1), having completed one period. Mathematically, this behavior can be well described by an asymptotic analysis in the continuum limit ($\nu \to 0, N\nu = const$) [32].

Close to the Hopf bifurcation described below we also find a different mode of oscillation. We no longer see recycling and motion of domain walls. Instead, a domain wall remains pinned at a given location and the field values at the low and high field domains oscillate in antiphase. The shape of the current oscillation is almost sinusoidal and the maximum current is clearly below 1 (where the accumulation layers were injected from the injecting contact for the mode discussed before). This mode has been experimentally observed in [31]. See Fig. 6. A similar oscillation mode has been found in simulations of imperfect superlattices [19].

Figures 7 show the variation in magnitude and frequency of the current oscillations as functions of the bias for different doping values and number of SL periods. In Fig. 7(a), the oscillatory behavior begins at $\phi = \phi_\alpha \approx 1.100$ via a supercritical Hopf bifurcation. The amplitude of the oscillation increases with bias following a square-root law and the frequency is almost constant. At $\phi = \phi_\omega \approx 1.617$ the branch of oscillatory solutions disappears via a second Hopf bifurcation. For longer SLs with $N = 200$ and the same doping, Fig. 7(e), or for a 50-well SL with larger doping, 7(c), the end of the oscillatory branch is different: a limit cycle collides with the unstable fixed point from the middle branch of the Z-shaped current-voltage characteristics and disappears. This bifurcation scenario results in a decrease of the frequency down till zero, when the collision takes place, while the amplitude is unchanged in lowest order. In all these examples, there is an interval of bistability for $\phi \in (\phi_\beta, \phi_\omega)$. The bias interval where the oscillatory branch exists, $(\phi_\alpha, \phi_\omega)$, shrinks as $\nu$ decreases, and for $\nu < \nu_0$ ($\nu_0 \approx 0.073$ for $N = 50$, $c = 10^{-4}$), there is no oscillatory solution.

It is important to note the dependence of the frequency with the bias, which could be used to tune the frequency of an eventual device [31]. The frequency might increase or decrease with bias depending on the sample parameters $N$ and $\nu$, as shown in Figs. 7(b,d,f).

In order to understand the different types of behavior we calculate the period $T_p$ of one oscillation. Let us start at the time $t = t_0$ when the the charge accumulation is injected at $i = 1$. One period is completed at $t = t_0 + T_p$ when the next charge accumulation is injected. At this instant, the position of the charge monopole is given by Eq. (31) with $E_l = 1$ and $E_h \approx E^{(3)}(1)$ which yields

$$M := j(t_0 + T_p) = \frac{E^{(3)}(1) - \phi}{E^{(3)}(1) - 1} N. \tag{33}$$

Thus the charge accumulation has to travel a distance $M$ in the time $T_p$. During the monopole formation time $T_f$, the mean velocity of the charge accumulation is $\nu^{-1}$, whereas it is equal to $v_{mon}/\nu$ (with $v_{mon} < 1$) for the rest of the period, $T_p - T_f$. We therefore have

$$M = \frac{1}{\nu} T_f + \frac{v_{mon}}{\nu} (T_p - T_f). \tag{34}$$

This gives

$$T_p = \frac{\nu M}{v_{mon}} - T_f \left( \frac{1}{v_{mon}} - 1 \right). \tag{35}$$

In order to estimate $T_f$, we note that is determined by the condition $\delta n(t_0 + T_f)/\delta n(t_0) = a$ where $a \sim 1/c$. Then Eq. (32) yields:

$$\log(a) = -\int_{t_0}^{t_0+T_f} v'(E_l(t))\, dt. \tag{36}$$

Now we obtain $E_l(t)$ from (31) with $E_h \approx E^{(3)}(1)$, $j(t) = M + v_{mon}(t - t_0)/\nu$. Up to the first order in $(t - t_0)$, this yields:

$$E_l(t) = 1 + \frac{N v_{mon}(E^{(3)}(1) - \phi)}{\nu M^2}(t - t_0). \tag{37}$$

Linearizing $v'(E_l) \approx -|v''(1)|(E_l - 1)$, we obtain

$$T_f = \sqrt{\nu M} \cdot \sqrt{\frac{2 \log(a)}{|v''(1)| v_{mon}(E^{(3)}(1) - 1)}}. \tag{38}$$

Now $M$ decreases as $\phi$ increases, as can be seen from Eq. (33). As $v_{mon} < 1$, Eq. (35) indicates that there are two competing mechanisms influencing the dependence of $T_p$ with $\phi$. In general the monopole formation time $T_f$ is negligible compared to $T_p(\phi)$ in the following cases: (i) for long SLs ($M$ in Eq. (35) is then large), (ii) when the values of $c$ are large (i.e., $a$ small), (iii) for large doping $\nu$. In these cases, $T_p(\phi)$ should be a decreasing function. We find that the frequency increases with $\phi$ for large values of the quantities $N$, $\nu$, $c$ while it decreases otherwise. These behaviors are illustrated in Fig. 8.

Until now we have studied Equations (6) and (7), with the boundary condition (8), $c$, positive. Now we are going to describe what happens if $-1 < c < 0$, that is, there are less electrons in the first well than the doping density ($n_1 - 1 = c < 0$; see Fig. 9).

As in the case of positive $c$, the oscillations are due to the generation, motion and annihilation of domain walls, connecting domains (which are regions of almost uniform electric field approximately given by the zeros of $I - v(E)$). The difference is that now the electric field profile is monotone decreasing with the QW index: the high-field domain is close to the beginning of the SL and the low-field domain extends to the end of the SL. The domain walls are now charge-depletion layers, having less electrons than the doping density. Let us describe one period of the current oscillations for a long SL, such as that in Fig. 9 with $N = 200$. We will assume that initially (point marked with 1 in Fig. 9(a)) the field profile has two domains connected by a domain wall ($E = E^{(3)}(I_0)$ to the left of the domain wall and $E = E^{(1)}(I_0)$ to the right of the domain wall, with an initial value of the current $I_0 \in (v_m, 1)$). The domain wall is approximately centered at $j/N = Y = (\phi - E^{(1)}(I_0))/(E^{(3)}(I_0) - E^{(1)}(I_0))$, so that Equation (7) holds. The domain wall then moves towards the end of the superlattice a with speed close to the instantaneous value of the current. The current decreases until a certain minimum value slightly smaller than $v_m$. Then a new high-field domain is created (close to the beginning of the SL) and the current rises sharply as the two domain walls move toward the end of the SL. When the current is near its maximum value, the old domain wall disappears as the values of the field on the intermediate and rightmost domains coalesce. Then one period of the oscillation is completed. All these features can be understood by means of an asymptotic analysis to be reported elsewhere [22].

## VI. CONCLUSIONS

In this paper we have shown how the complex stationary current-voltage characteristic exhibiting domain branches is generated continuously as the doping increases. For low doping the characteristic follows the local $v(E)$ relation. If more charges are present, the characteristic becomes Z-shaped. When the doping is even larger, wiggles appear. For each doping the characteristic is connected, and the field profiles of all its different branches are monotone. The different disconnected branches observed experimentally correspond to the stable solution branches of the full stationary current-voltage characteristic. It would be very interesting to investigate whether it is possible to stabilize the unstable branches so that the full characteristic could be observed, as in the case of the double-barrier resonant-tunneling diode [33,27]. Additionally, for large doping there exist isolated branches (isolas) on the full current-voltage characteristic having non-monotonic field profiles.

The stability analysis shows that the almost uniform field profile is stable for low doping. The critical doping above which time-periodic oscillations of the current appear is inversely proportional to the sample length for fixed superlattice parameters. This is the same situation as in the famous $N_D L$ criterion for the Gunn Diode. For yet larger doping the time-periodic oscillations of the current disappear: there is an upper critical doping above which there appear stable stationary solutions with two electric field domains (separated by an abrupt domain wall extending almost one period of the superlattice). Obviously, this is not possible for the conventional Gunn Diode due to the lack of discretization. It is important to mention that the upper critical doping needed to stabilize stationary domain structures is higher for profiles having depletion layers instead of accumulation layers between the different domains.

The transition from oscillatory behavior to a stationary state may occur via two different scenarios. For certain parameter values, we have found that the stationary and oscillatory solution branches coalesce by means of a supercritical Hopf bifurcation. In this case the amplitude of the oscillations drops to zero continuously at an almost constant frequency. On the other hand, the oscillatory solution may collide with the homoclinic orbit of a saddle point. In this case the oscillation frequency drops to zero while the amplitude remains finite and it does not change very much. The latter case has also been found numerically for different parameter values and it resembles the behavior found experimentally under illumination [10]. Depending on the charge at the first QW, the doping and the shape of the velocity curve, the frequency of the oscillation may depend in different ways on the dc voltage bias. This is also corroborated by experiments [10,11,31].

## ACKNOWLEDGMENTS

We thank J. Galán, H. T. Grahn, J. Kastrup, M. Patra, F. Prengel, G. Schwarz, E. Schöll and S. Venakides for fruitful discussions and collaboration on related topics. We thank E. Doedel for sending us his program of numerical continuation AUTO. This work has been supported by the DGICYT grants PB92-0248 and PB94-0375, and by the EU Human Capital and Mobility Programme contract ERBCHRXCT930413.

## APPENDIX A: PROOF OF STABILITY FOR SUFFICIENTLY SMALL $\nu$

Here we prove that $\lambda = i\omega$ with $\omega > 0$ can be an eigenvalue of the linearized system (19) only if $\nu$ is sufficiently large.

In order to do this, we assume that a given stationary field profile $\{E_i^c\}$ exhibits $\lambda = i\omega$ with $\omega > 0$ and derive several necessary conditions for this. Restricting ourselves to monotone increasing field profiles, $E_1^c$ must either be located on the first or second branch of the $v(E)$ curve, as otherwise the second branch is not reached which is a necessary condition for the instability according to Lemma 3. Let us now determine the smallest value $E_l$ and the largest value $E_h$ the stationary field profile $\{E_i^c\}$ may take. $E_l$ is given by the value of $E_1^c$ on the first branch for which the current takes on its minimal value, $I^* = 1$, considering that the field must eventually take values on the second branch of $v(E)$. Eq. (14) yields $v(E_l) = 1/(1+c)$ which determines the field $E_l \leq 1$. $E_h$ is given by the largest value that $E_N^c$ can take on the third branch of $v(E)$. Noticing that $I^* \leq 1 + c$ in Eq. (14), we can adopt $E_h$ as the solution of $v(E_h) = 1 + c$ from the third branch. Thus, $E_l$ and $E_h$ depend only on the $v(E)$ curve and the parameter $c$ but not on the field profile $\{E_i^c\}$. For sake of convenience we introduce the following quantities:

$$v_l := \min\left\{v_m, \frac{1}{1+c}\right\} \tag{A1}$$

$$c_1 := I^* \max_{E_l \le E \le E_h} \left|\frac{\partial \ln v(E)}{\partial E}\right| \tag{A2}$$

$$c_2 := I^* \max_{E_l \le E \le E_h} \left|\frac{\partial^2 \ln v(E)}{\partial E^2}\right| \tag{A3}$$

Then we have

$$a_i \ge \frac{v_l}{\nu} \quad \text{and} \quad |b_i| \le c_1 \quad \forall i \tag{A4}$$

In the following we will assume that $\nu$ is so small that the function $f(E, I^*)$ ($I^*$ fixed) is always invertible and furthermore

$$a_i + b_i > v_l/\nu - c_1 > 0\,, \quad \forall i \tag{A5}$$

holds.

For $\lambda = i\omega$, $\omega > 0$, we have $\mathrm{Re}Y_1 > 0$, and $|Y_1 - 1| < 1$. As previously explained in the proof of Lemma 3, there must be a $Y_j$ such that $\mathrm{Re}Y_i \ge 0$, $i = 1, \ldots, j-1$, and $\mathrm{Re}Y_j < 0$ in order to fulfil the voltage condition. We are going to prove the following result:

**Lemma 4** *Let $j > 1$ be the index that satisfies $ReY_i \ge 0$, $i = 1, \ldots, j-1$, and $ReY_j < 0$.*
*(a) If $\omega \le c_1$, we have*

$$(j-1)\,\nu > \frac{\pi v_l}{4Ac_1} \tag{A6}$$

*where $A$ is the maximum of the expressions*

$$A_k := \frac{v_l}{(j-k+1)\nu} \sum_{i=k}^{j} \frac{1}{a_i + b_i}\,, \tag{A7}$$

*for $k = 2, \ldots j$.*
*(b) If $\omega > c_1$, we have*

$$(j-1)\,\nu > \frac{v_l c_1}{2B\,(I^* - v_l)\,c_2}\,, \tag{A8}$$

*where*

$$B := \frac{\mu\,(\mu^{j-1} - 1)}{(\mu - 1)\,(j-1)}\,, \quad \textit{with} \tag{A9}$$

$$\mu := max_{E_l \le E_i \le E_h} \left\{\frac{a_i}{|a_i + b_i + ic_1|}\right\}. \tag{A10}$$

*Proof*:
<u>**(a)** Let $\omega \le c_1$:</u>
In order to prove (A6), we consider how the argument $\phi_i$ of the complex quantity $Y_i$ is varying with $i$.

$$\frac{Y_{i-1}}{Y_i} = \frac{|Y_{i-1}|}{|Y_i|} e^{i(\phi_{i-1} - \phi_i)} = \frac{1}{a_i}\left(i\omega(1 - Y_i^{-1}) + b_i + a_i\right) \tag{A11}$$

Therefore we get:

$$\phi_{i-1} - \phi_i = \arctan\left(\frac{\omega - \omega Re(Y_i^{-1})}{b_i + a_i + \omega Im(Y_i^{-1})}\right) \tag{A12}$$

Furthermore we have:

$$\phi_1 = \arctan\left(\frac{b_1}{\omega}\right) \tag{A13}$$

Straightforward calculations starting from eq. (22) yield:

$$Re(Y_i) = \frac{\omega^2 + a_i\omega Im(Y_{i-1}) + a_i(b_i + a_i)Re(Y_{i-1})}{\omega^2 + (b_i + a_i)^2}$$
$$Re(Y_i) = \frac{\omega}{b_i + a_i} Im(Y_i) + \frac{a_i}{b_i + a_i} Re(Y_{i-1}) \tag{A14}$$

By definition of the index $j$ ($j \le N$), $\mathrm{Re}Y_j < 0$ and $\mathrm{Re}Y_{j-1} \ge 0$. Then these equations indicate that $\mathrm{Im}Y_j < 0$ and $\mathrm{Im}Y_{j-1} < 0$. Thus the transition $\mathrm{Re}Y_{j-1} \ge 0 \to \mathrm{Re}Y_j < 0$ occurs across the angle $\phi = -\pi/2$ as we have $-\pi/2 \le \phi_{j-1} < 0$ and $-\pi < \phi_j < -\pi/2$.

We introduce the index $j'$ which is defined by the relations $-\pi/4 \le \phi_{j'-1}$ and $\phi_i < -\pi/4$ for $i = j', \ldots, j$. Obviously, we have $\mathrm{Im}Y_i < 0$ and therefore $b_i + a_i - \omega\frac{Im(Y_i)}{|Y_i|^2} > 0$ for $i = j', j'+1, \ldots, j$. Using (A14) and $\mathrm{Re}Y_{i-1} \ge 0$ (for all $i \le j$), we obtain $\mathrm{Re}Y_i \ge \omega$ $\mathrm{Im}Y_i/(a_i + b_i)$. This yields for $i = j', \ldots, j$:

$$\begin{aligned}\phi_{i-1} - \phi_i &= \arctan\left(\frac{\omega - \omega\frac{Re(Y_i)}{|Y_i|^2}}{b_i + a_i - \omega\frac{Im(Y_i)}{|Y_i|^2}}\right) \\ &\le \arctan\left(\frac{\omega - \frac{\omega^2 Im(Y_i)}{(b_i+a_i)|Y_i|^2}}{b_i + a_i - \omega\frac{Im(Y_i)}{|Y_i|^2}}\right) \\ &= \arctan\left(\frac{\omega}{a_i + b_i}\right) < \frac{\omega}{a_i + b_i}\end{aligned} \tag{A15}$$

Now we have to distinguish two different cases:
i) $j' \ge 2$: By summing the inequality (A15) from $i = j'$ to $i = j$ and then taking into account the definitions of $j$ and $j'$, we find:

$$\begin{aligned}-\frac{\pi}{4} + \frac{\pi}{2} < \phi_{j'-1} - \phi_j &< \sum_{i=j'}^{j} \frac{\omega}{a_i + b_i} \\ &= \frac{\omega\,(j - j' + 1)\,\nu\,A_{j'}}{v_l}\,,\end{aligned} \tag{A16}$$

where definition (A7) has been used. The property $\omega \le c_1$ then implies

$$j-1 \geq j-j'+1 > \frac{\pi v_l}{4A_{j'}\nu c_1}. \quad \text{(A17)}$$

ii) $j'=1$: This means that $\phi_1 < -\pi/4$ and according to to Eq. (A13), $\omega < -b_1 = |b_1|$. Now we sum the inequality (A15) from $i=2$ to $i=j$ and then use the expression $\arctan x > \pi x/4$ for $0<x<1$, thereby obtaining

$$\frac{\pi\omega}{-4b_1} < \arctan\left(\frac{\omega}{-b_1}\right) = \arctan\left(\frac{b_1}{\omega}\right) + \frac{\pi}{2}$$
$$< \phi_1 - \phi_j < \frac{\omega\,\nu\,(j-1)\,A_2}{v_l}. \quad \text{(A18)}$$

Therefore we find with Eq. (A4):

$$j-1 > \frac{\pi v_l}{4\,\nu\,A_2\,|b_1|} \geq \frac{\pi v_l}{4\,A_2\,\nu c_1}. \quad \text{(A19)}$$

Putting together (A17) and (A19) we obtain the inequality (A6).

**(b)** Let now $\omega > c_1$:

To prove the inequality (A8), we shall define the auxiliary functions

$$Z_i = Y_i - \frac{\lambda}{\lambda + b_i}. \quad \text{(A20)}$$

These functions solve the following discrete equation

$$(\lambda + a_i + b_i)\,Z_i - a_i\,Z_{i-1} = \lambda a_i\,\left(\frac{1}{\lambda + b_{i-1}} - \frac{1}{\lambda + b_i}\right), \quad \text{(A21)}$$

with the boundary condition $Z_1 = 0$. The solution of this problem is

$$Z_n = \sum_{k=2}^{n} \frac{\lambda\,(b_k - b_{k-1})}{(\lambda + b_{k-1})\,(\lambda + b_k)} \prod_{i=k}^{n} \frac{a_i}{\lambda + b_i + a_i}. \quad \text{(A22)}$$

As all $b_i$ are real quantities, we have $|i\omega + b_i| > \omega$ and obtain the following inequality for $|Z_j|$ by using the preceding formula with $\lambda = i\omega$, $\omega > 0$:

$$|Z_j| < \frac{1}{\omega} \sum_{k=2}^{j} |b_k - b_{k-1}| \prod_{i=k}^{j} \frac{a_i}{|b_i + a_i + i\omega|}. \quad \text{(A23)}$$

Now we have $|Z_j| > -\,\mathrm{Re}Z_j = -\,\mathrm{Re}Y_j + \omega^2/(\omega^2 + b_j^2) > \omega^2/(\omega^2 + b_j^2)$, where $\mathrm{Re}Y_j < 0$ and the definition of $Z_n$ have been used. This inequality together with (A23) yield

$$\frac{\omega^3}{\omega^2 + b_j^2} < \sum_{k=2}^{j} |b_k - b_{k-1}| \prod_{i=k}^{j} \frac{a_i}{|b_i + a_i + i\omega|}. \quad \text{(A24)}$$

We now estimate the right side of (A24). The definition (21) of $b_i$ and the mean value theorem yield

$$|b_k - b_{k-1}| < c_2\,|E_k^c - E_{k-1}^c|. \quad \text{(A25)}$$

Equation (13) for the stationary state now yields $0 < E_k^c - E_{k-1}^c = (I^*/a_k - \nu)$, so that $0 < E_k^c - E_{k-1}^c < \nu(I^*/v_l - 1)$. Thus we can write:

$$|b_k - b_{k-1}| < \nu\left(\frac{I^*}{v_l} - 1\right)\,c_2. \quad \text{(A26)}$$

On the other hand, as we are considering the case $\omega > c_1$, we find that

$$\frac{a_i}{|b_i + a_i + i\omega|} < \frac{a_i}{|b_i + a_i + ic_1|} \leq \mu, \quad \text{(A27)}$$

according to (A10). Inserting (A26) and (A27) into (A24), we obtain

$$\frac{\omega^3}{\omega^2 + c_1^2} < \nu\,c_2\left(\frac{I^*}{v_l} - 1\right)\left(\sum_{k=2}^{j} \mu^{j-k+1}\right)$$
$$= \nu\,c_2\left(\frac{I^*}{v_l} - 1\right)\mu\,\frac{\mu^{j-1} - 1}{\mu - 1}. \quad \text{(A28)}$$

Since $\omega > c_1$, $\omega^3/(\omega^2 + c_1^2) > c_1/2$. Inserting this into (A28), we obtain (A8). Therefore Lemma 4 is proved.•

Lemma 4 yields necessary conditions for the instability of a given stationary field profile $\{E_i^c\}$ corresponding to a fixed bias. We would like to obtain a general condition on $\nu$, which should only depend on the $v(E)$ curve and the parameter $c$, but not on the specific stationary field profile. This can be achieved by the following considerations:

$$A_k = \frac{v_l}{(j-k+1)\nu} \sum_{i=k}^{j} \frac{1}{a_i + b_i} \leq \frac{v_l}{v_l - \nu c_1}. \quad \text{(A29)}$$

Therefore we have $A < v_l/(v_l - \nu c_1)$ from Eq. (A7), which inserted into the inequality (A6) gives

$$(j-1)\nu > \pi\frac{v_l - \nu c_1}{4c_1}. \quad \text{(A30)}$$

From the definition (A10) we obtain

$$\mu \leq \max_{E_l \leq E_i \leq E_h}\left\{\frac{a_i}{a_i - c_1}\right\} \leq 1 + \frac{\nu c_1}{v_l - \nu c_1}. \quad \text{(A31)}$$

This yields

$$B \leq \frac{v_l}{(j-1)\nu c_1}\left[\left(1 + \frac{\nu c_1}{v_l - \nu c_1}\right)^{j-1} - 1\right]$$
$$\leq \frac{v_l}{(j-1)\nu c_1}\left[\exp\left(\frac{(j-1)\nu c_1}{v_l - \nu c_1}\right) - 1\right] =: C, \quad \text{(A32)}$$

to be inserted in (A8). The result is

$$(j-1)\,\nu > \frac{v_l c_1}{2C\,(I^* - v_l)\,c_2}, \quad \text{(A33)}$$

We now use the obvious inequality $N \geq j$ in (A30) and (A33) thereby obtaining the condition (30) as a necessary condition for oscillatory instability of the steady state.

---

* Now at: Mikroelektronik Centret, Technical University of Denmark, DK-2800 Lyngby, Denmark

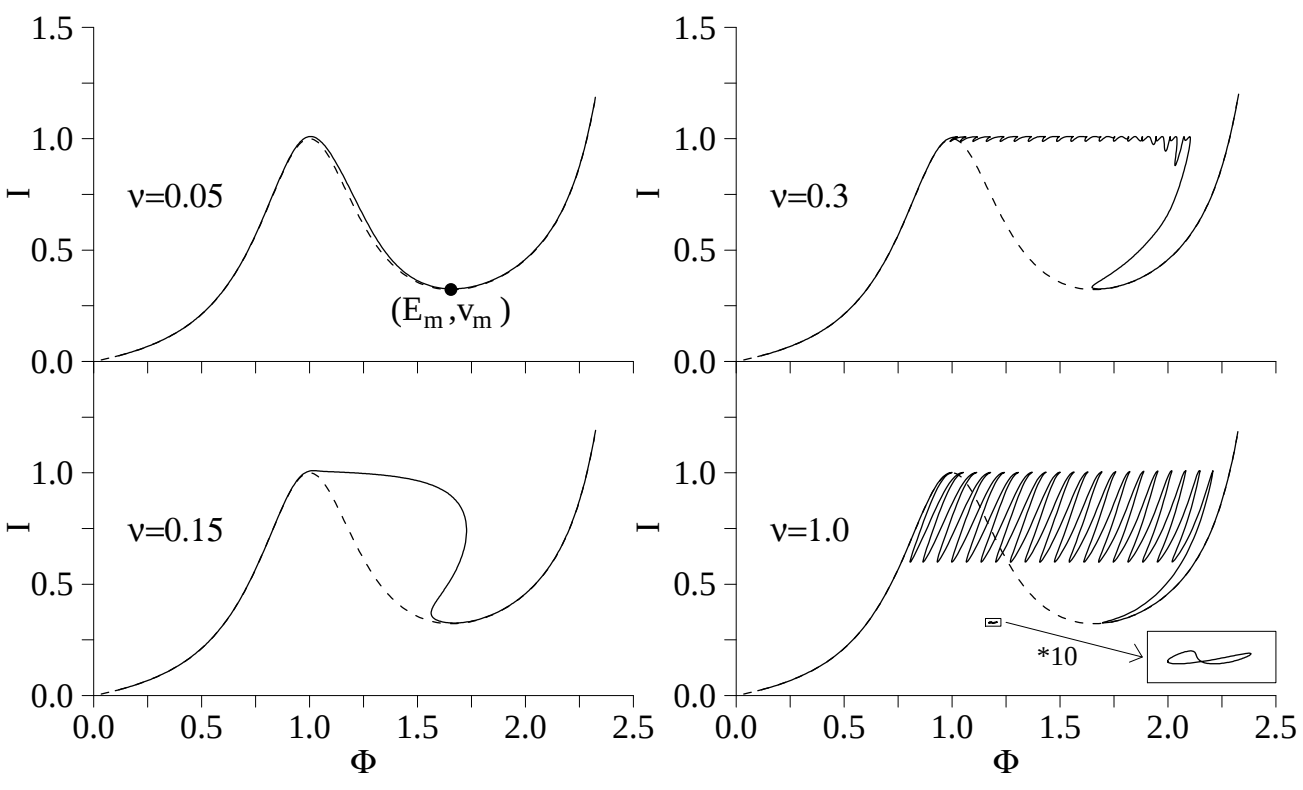

FIG. 1. Current-voltage characteristics for $c = 0.01$ and $N = 20$ and different values of $\nu$. The full line denotes the states where the electric field $E_i$ is strictly monotone increasing in $i$. For $\nu = 1.0$ there appear additional branches with non-monotonic field profiles $E_i$. They are isolated from the stationary branches having monotone increasing field profiles. We have shown one such branch, which has also been blown up for the sake of clarity. The dotted line is the $v(E)$ curve used throughout this paper.

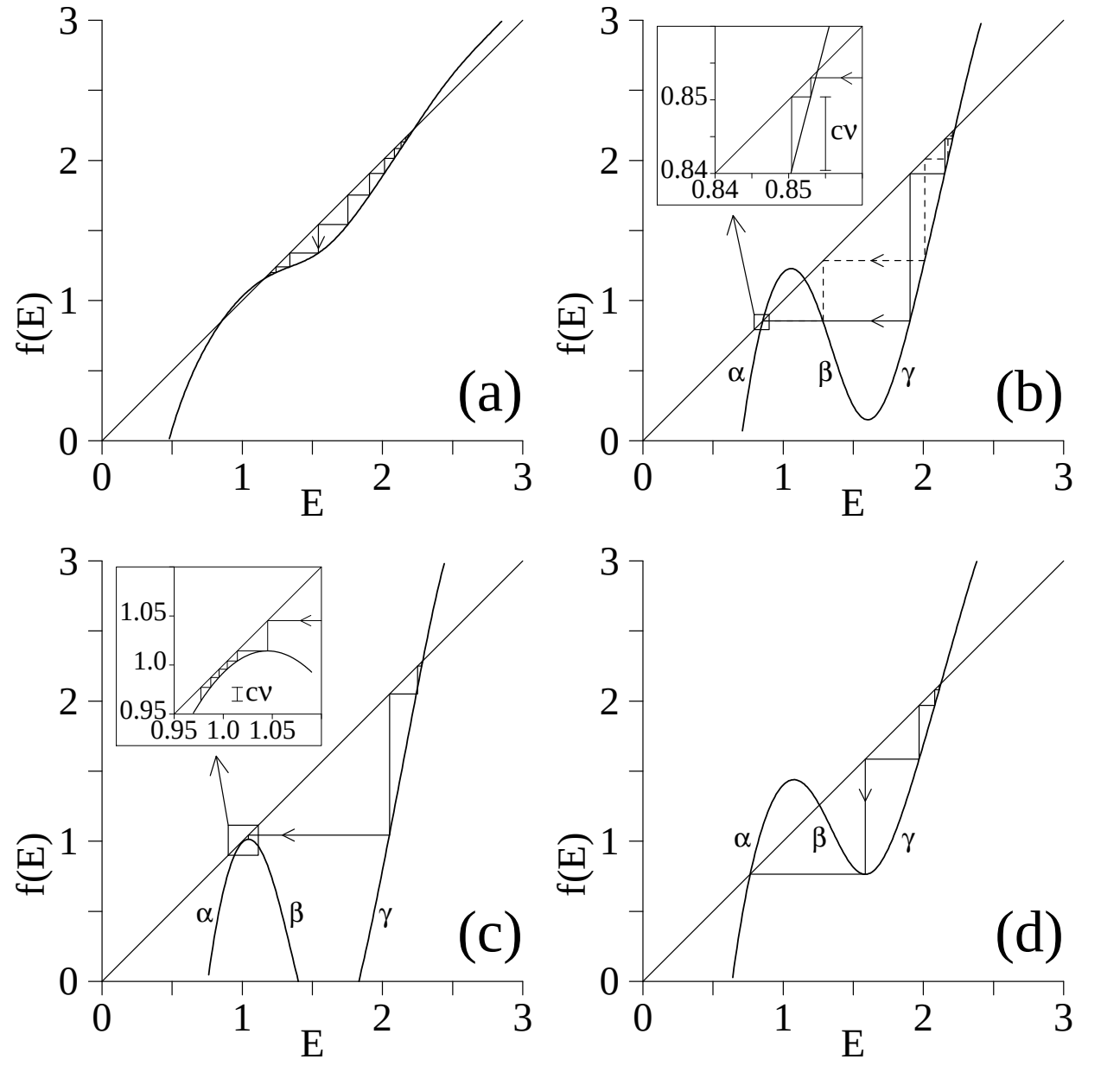

FIG. 2. $f(E, I)$ and a trajectory $E_i$ indicating decreasing $i$ for various doping densities and currents. (a) $\nu = 0.15$, $I = 0.8$. (b) $\nu = 1.0$, $I = 0.8$. (c) $\nu = 1.0$, $I = 1.008$. (d) $\nu = 1.0$, $I = 0.599 = I_c$.

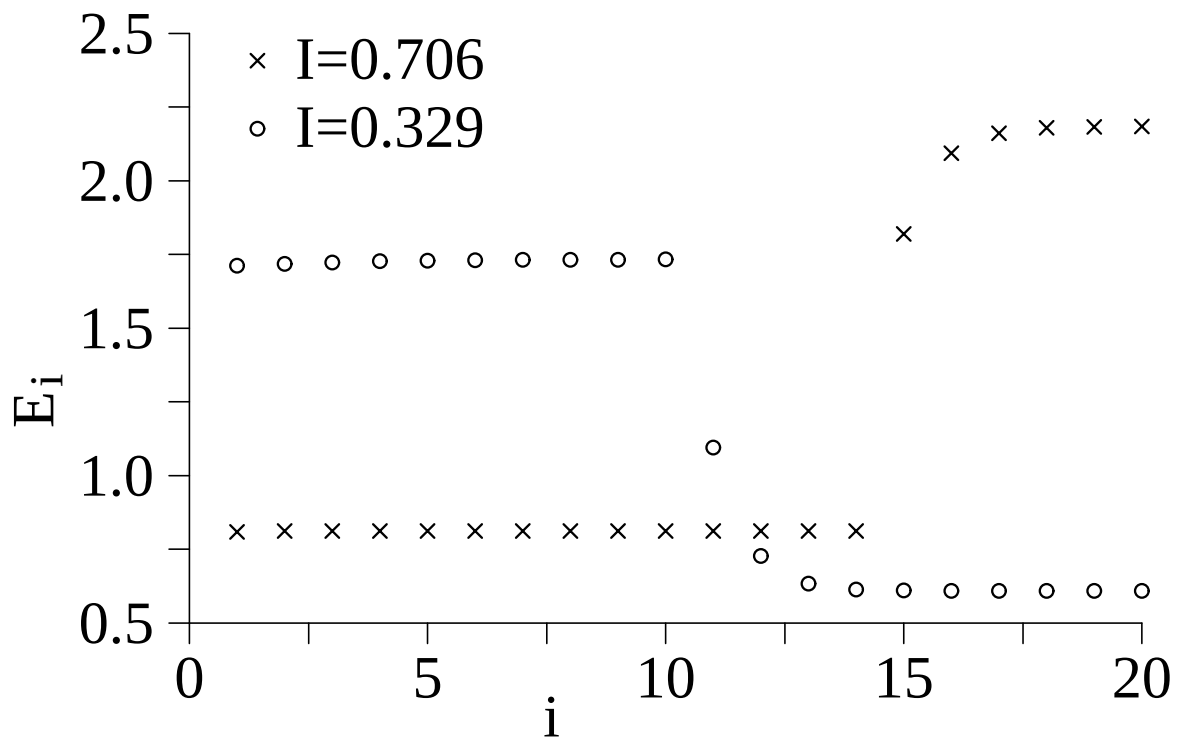

FIG. 3. Different stationary electric field profiles for $c = 0.01$, $\nu = 1.0$, $\phi = 1.2$, and $N = 20$. The crosses mark a state of the connected branch from Fig. 1($\nu = 1.0$) while the circles mark a state belonging to the isolated branch.

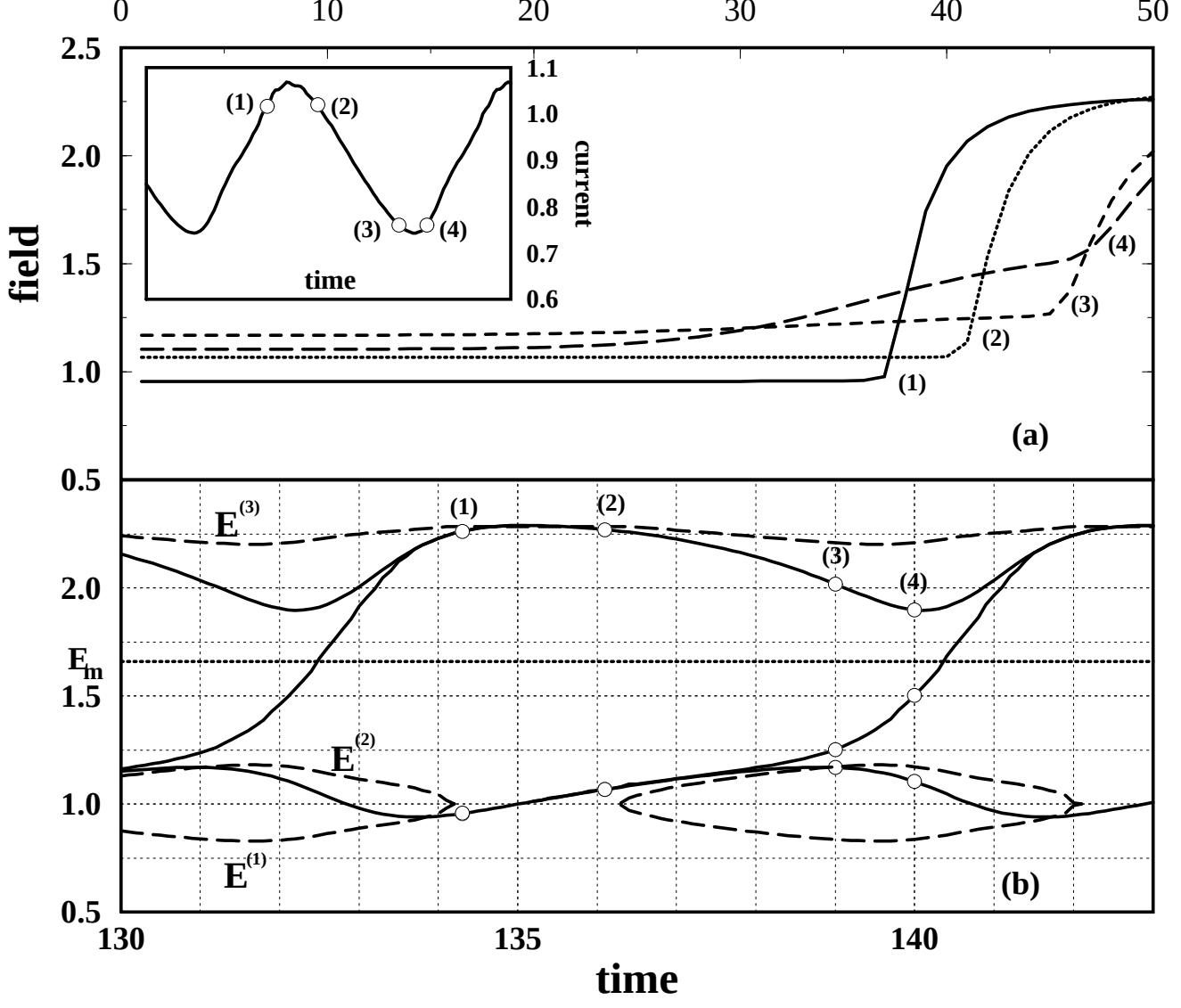

FIG. 4. (a) Electric field profiles at different times during one period of the current oscillation depicted in the inset. (b) Time evolution of the electric field values in the left, middle (when it exists) and right domain of the SL. The corresponding values of $E^{(i)}(I(t))$, $i = 1, 2, 3$, are represented with dashed lines. Parameter values are $\phi = 1.25$, $c = 10^{-4}$, $\nu = 0.1$, and $N = 50$.

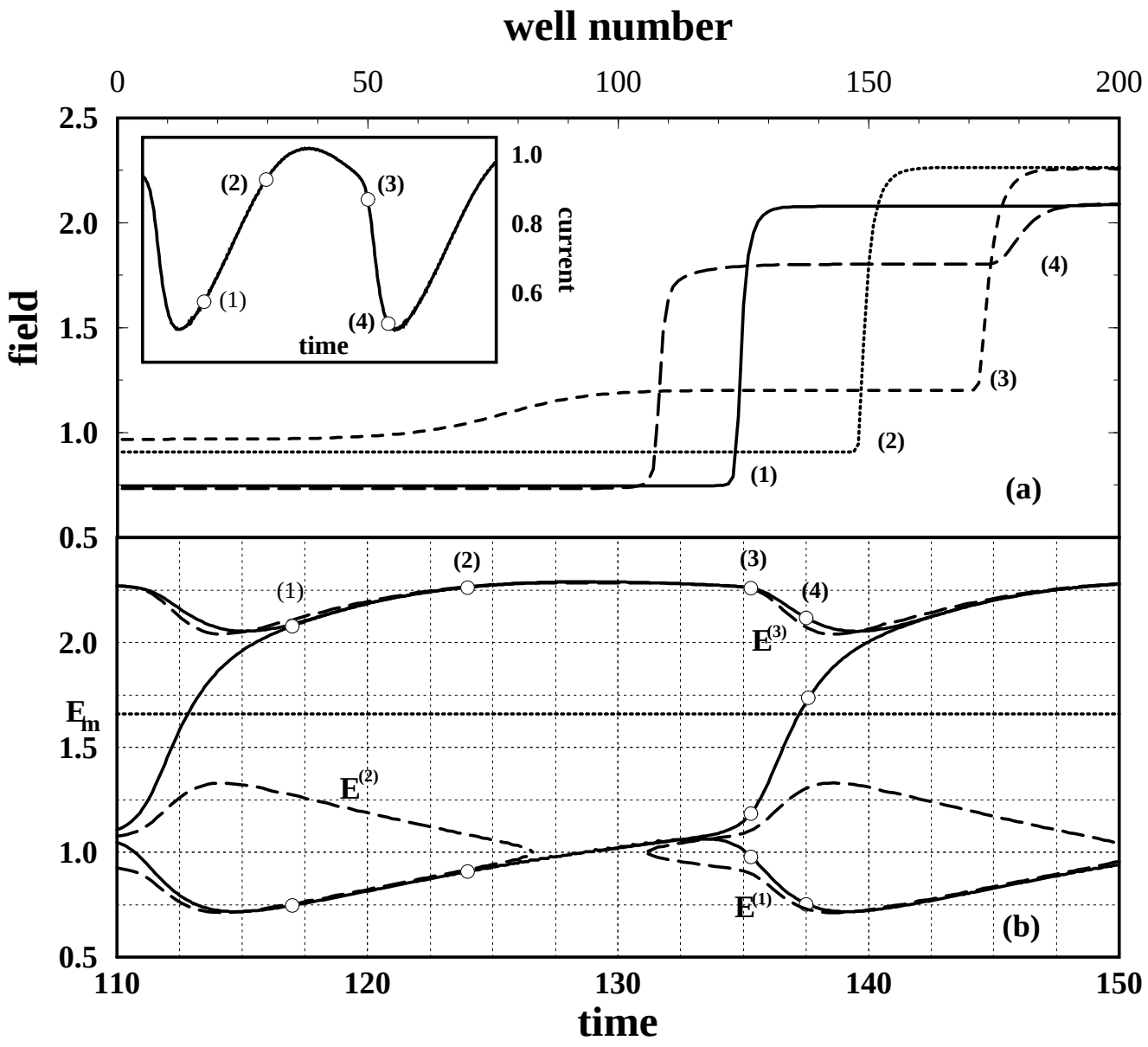

FIG. 5. Same as in Figure 4 for $N = 200$.

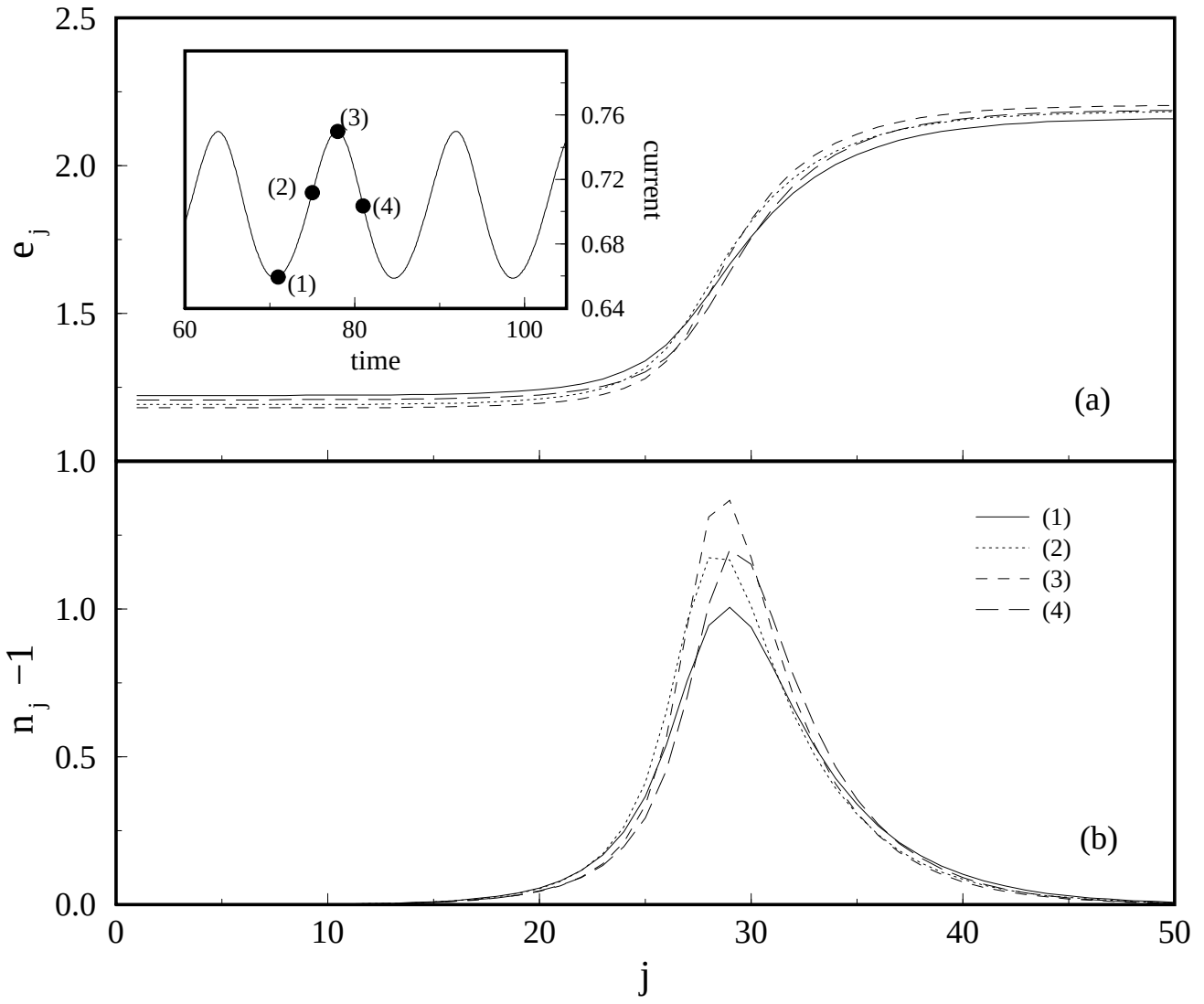

FIG. 6. (a) Same as Fig. 4(a) for N=50 and $\phi = 1.61$ close to the Hopf bifurcation. (b) Charge density profiles corresponding to the electric field profiles shown in (a).

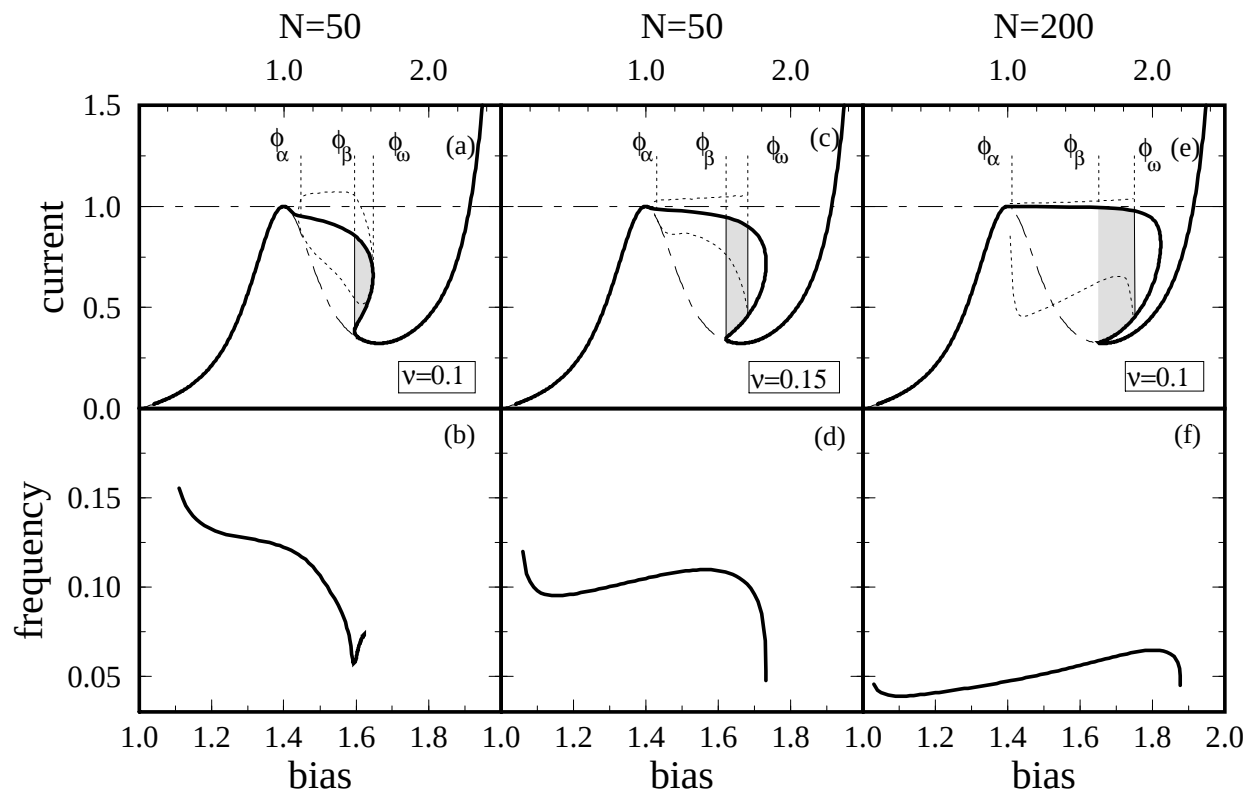

FIG. 7. (a) Stationary current-voltage characteristic (full line) with maximum and minimum of the oscillating current (dotted lines) $N = 50$. The oscillatory branch begins at $\phi_\alpha \approx 1.100$ and ends at $\phi_\omega \approx 1.617$; the interval of bistability begins at $\phi_b \approx 1.487$. (b) Fundamental frequency of the current vs. the average electric field (bias) for $N = 50$. (c,d) Same as (a,b) but now $\nu = 0.15$; $\phi_\alpha \approx 1.052$, $\phi_b \approx 1.556$, $\phi_\omega \approx 1.797$. (e,f) Same as (a,b) but $N = 200$; $\phi_\alpha \approx 1.02$, $\phi_b \approx 1.622$, $\phi_\omega \approx 1.868$. In all cases $c = 10^{-4}$.

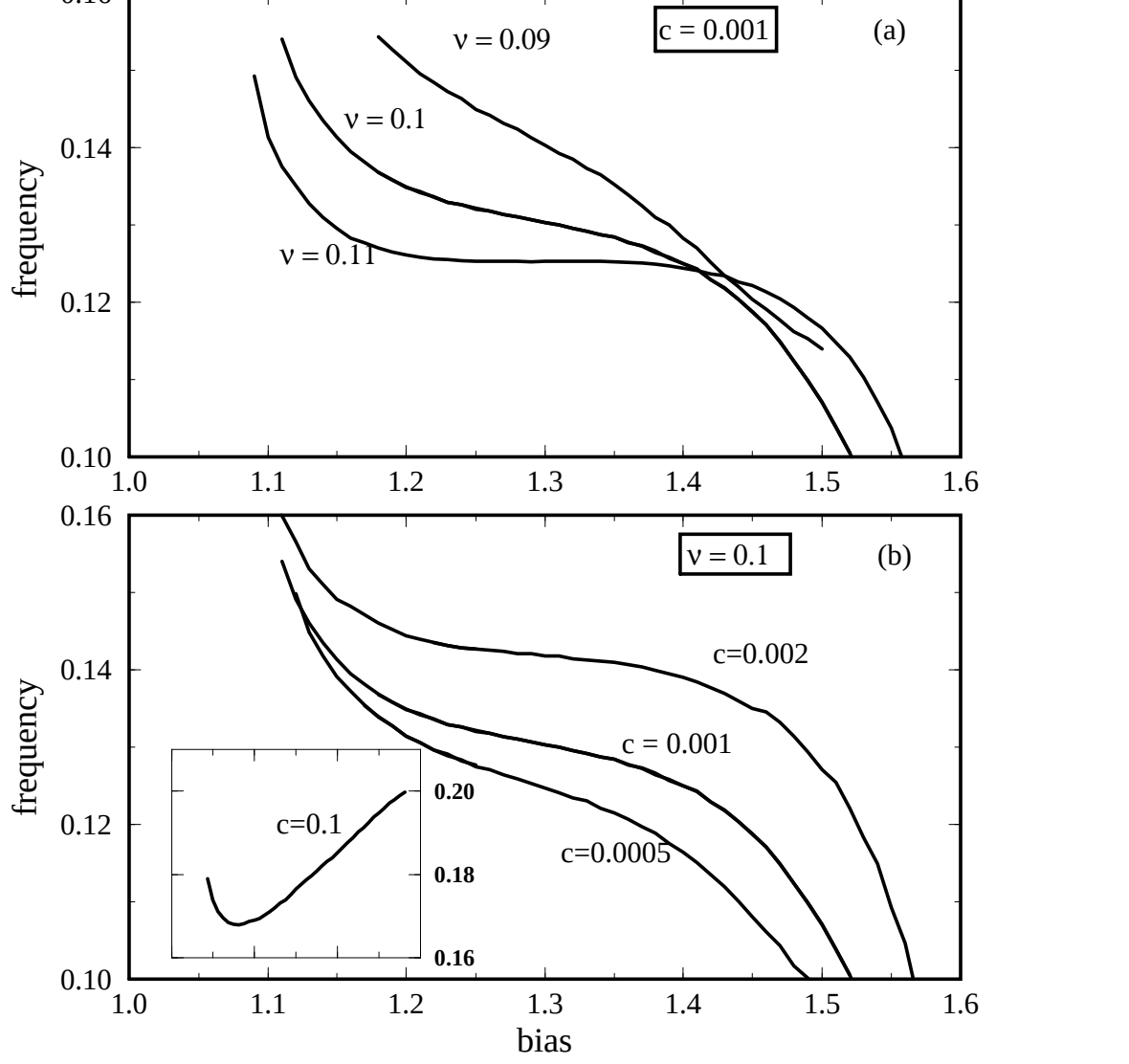

FIG. 8. (a) Frequency vs. bias diagrams for different values of the dimensionless doping $\nu$ and $c = 0.001$. (b) Frequency vs. bias diagrams for different values of $c$ and $\nu = 0.1$. $N = 50$.

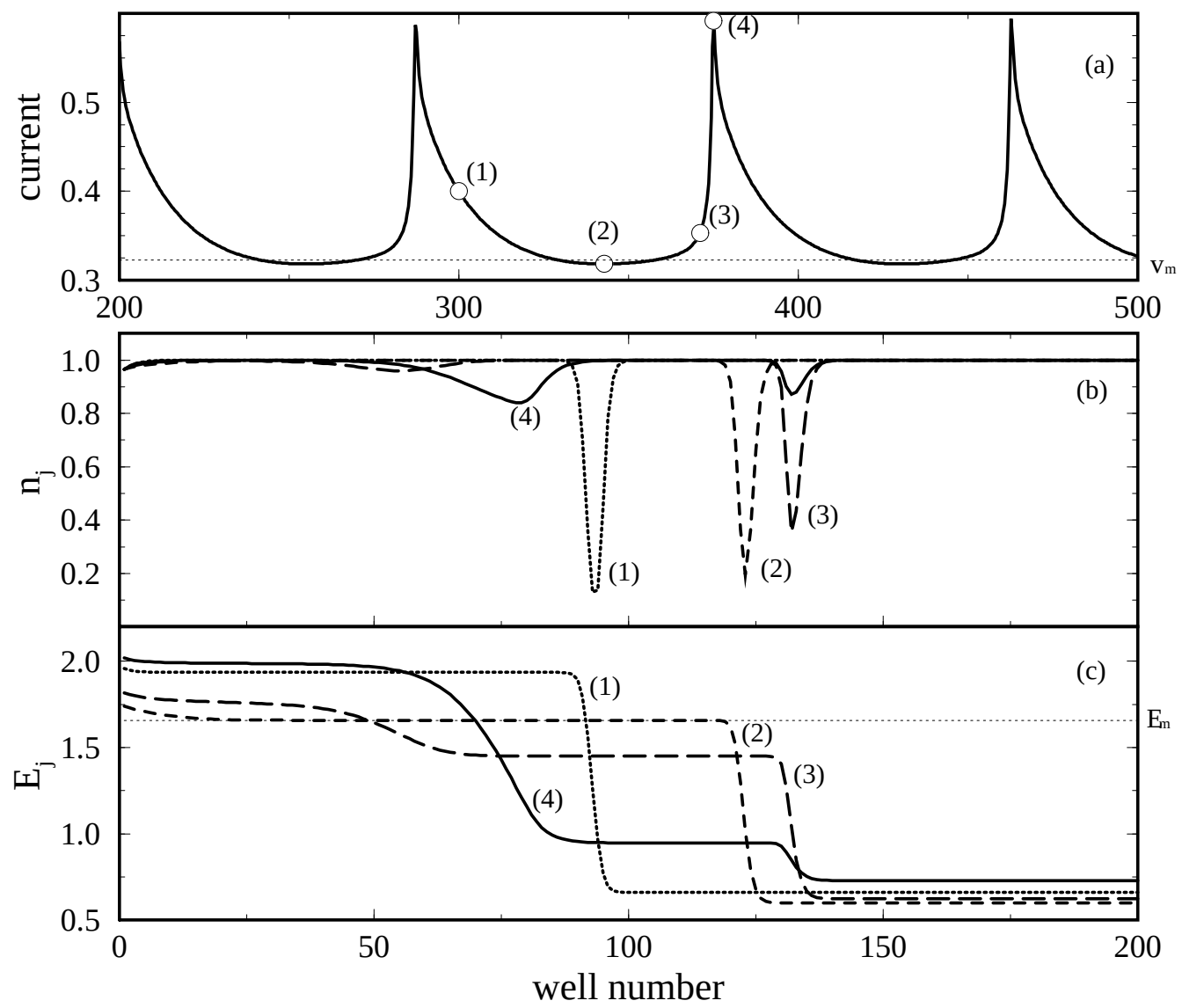

FIG. 9. (a) Current density versus time when $c = -0.01$. Charge density (b) and electric field (c) profiles during one period of the current oscillation. The numbers in (b) and (c) correspond to the times marked in (a). Parameter values are $\phi = 1.25$, $\nu = 0.35$, and $N = 200$.